\documentstyle[emulateapj5,epsf]{aastex}
\def \vt{\vartheta}
\def \farcs{\hbox{$.\!\!^{\prime\prime}$}}

\lefthead{Hoekstra et al.}
\righthead{Cosmological parameters from RCS}
\begin{document}

\title{Constraints on $\Omega_m$ and $\sigma_8$ from weak lensing in 
RCS fields}

\author{Henk~Hoekstra$^{1,2,3}$, H.K.C.~Yee$^{2,3,4}$, and Michael
D.~Gladders$^{2,3,4,5}$}

\begin{abstract}
We have analysed 53 square degrees of $R_C$-band imaging data from the
Red-Sequence Cluster Survey (RCS), and measured the excess
correlations in the shapes of galaxies on scales out to $\sim 1.5$
degrees. We separate the signal into an ``E''- (lensing) and
``B''-mode (systematics), which allows us to study the contribution
from residual systematics and intrinsic alignments.  On scales larger
than 10 arcminutes, we find no ``B''-mode, suggesting that the signal
at those scales is solely caused by gravitational lensing. On smaller
scales we find a small, but significant ``B''-mode. This signal is
also present when we select a sample of bright $(20<R_C<22)$
galaxies. These galaxies are rather insensitive to observational
distortions, and we therefore conclude that the observed ``B''-mode is
likely to be caused by intrinsic alignments. To minimize the effect of
intrinsic alignments, we limit the cosmic shear analysis to galaxies
with $22<R_C<24$.

We derive joint constraints on $\Omega_m$ and $\sigma_8$, by
marginalizing over $\Gamma$, $\Omega_\Lambda$ and the source redshift
distribution, using different priors. Marginalizing over $\Gamma$ and
$\Omega_\Lambda$, and using a flat prior for the source redshift
distribution, yields a conservative constraint of
$\sigma_8=0.45^{+0.09}_{-0.12}\Omega_m^{-0.55}$ (95\% confidence).  A
better constraint is derived when we use Gaussian priors for $\Gamma$
(from the 2dF survey) and $\Omega_m+\Omega_\Lambda$ (from CMB), and
the source redshift distribution. For this choice of priors, we find
$\sigma_8=(0.46^{+0.05}_{-0.07})\Omega_m^{-0.52}$ (95\% confidence).
We also investigated whether the RCS data can be used to constrain
$\Gamma$. Using our set of Gaussian priors, we find that we can only
place a lower bound on $\Gamma$ for which we find
$\Gamma>0.1+0.16\Omega_m$ (95\% confidence).

Comparison of the RCS results with three other recent cosmic shear
measurements shows excellent agreement. The current weak lensing
results are also in good agreement with CMB measurements, when we
allow the reionization optical depth $\tau$ and the spectral index
$n_s$ to vary. We present a simple demonstration of how the weak
lensing results can be used as a prior in the parameter estimation
from CMB measurements to derive constraints on the reionization
optical depth $\tau$. 

\end{abstract}

\keywords{cosmology: observations $-$ dark matter $-$ gravitational lensing}

\section{Introduction}

The measurement of the coherent distortions of the images of faint
galaxies caused by weak lensing by intervening large scale structures
provides a direct way to study the statistical properties of the
growth of structures in the universe. Compared to many other methods,
such as redshift surveys (e.g., Strauss \& Willick 1995; Peacock et
al. 2001; Percival et al. 2001), weak lensing probes the dark matter
distribution directly, regardless of the light distribution. In
addition, it provides measurements on scales from the quasi-linear to
the non-linear regime, where comparisons between observations and
predictions are still limited.

The distortion in the images of distant galaxies is small,
and consequently it is not measured easily. Only recently it has
become possible to measure the signal and the 

\vbox{
\vspace{0.5cm}
\footnotesize
\noindent 
$^1$~CITA, University of Toronto, Toronto, Ontario M5S 3H8, Canada\\
$^2$~Department of Astronomy, University of Toronto,
Toronto, Ontario M5S 3H8, Canada\\
$^{3}$~Visiting Astronomer, Canada-France-Hawaii Telescope, which
is operated by the National Research Council of Canada, Le Centre 
National de Recherche Scientifique, and the University of Hawaii\\
$^{4}$~Guest observer at the Cerro Tololo Inter-American Observatory
(CTIO), a division of the National Optical Astronomy Observatories,
which is operated by the Association of Universities for Research in
Astronomy, Inc., under cooperative agreement with the National Science
Foundation\\ 
$^{5}$~Observatories of the Carnegie Institution of 
Washington, 813 Santa Barbara Street, Pasadena, California 91101}

\noindent number of published measurements is increasing rapidly
(e.g., Bacon et al. 2000, 2002; Hoekstra et al. 2002a; Kaiser, Wilson,
\& Luppino 2000; Maoli et al. 2000; Refregier, Rhodes, \& Groth 2002;
van Waerbeke 2000, 2001a, 2002; Wittman et al. 2000).

Weak lensing observations can be used for a wide range of studies. For
instance, Hoekstra, Yee, \& Gladders (2001) presented a first
measurement of a combination of the bias parameter and the galaxy-mass
cross-correlation coefficient as a function of scale. The first direct
measurement of the bias parameter and the galaxy-mass
cross-correlation coefficient was presented by Hoekstra et
al. (2002b). These studies provide unique constraints on models of
galaxy formation.  Another important application is to constrain
cosmological parameters from the measurement of the two-point
statistics of the dark matter distribution. 

The amplitude of the lensing signal depends on many parameters, of
which some are degenerate. Some of the degeneracies can be overcome
using measurements that probe the dark matter power spectrum at
different redshifts, or by comparison of the results in the linear
regime to the non-linear scales. Such subtle differences require very
accurate measurements, which are currently not available.  Planned
large surveys, such as the CFHT Legacy Survey, however, will be a
major advance in this area. Also higher order statistics, such as the
skewness, are potentially powerful tools to break some of the
degeneracies (e.g., Schneider et al. 1998).

Better constraints can be obtained by combining the 
weak lensing measurements with results from other techniques.
In particular the combination with cosmic microwave background (CMB)
measurements is useful, because the physics of both lensing and
CMB are well understood. Also some results  from galaxy redshift
surveys can be used, provided they are made in the linear regime,
and under the assumption that the bias is constant with scale.

In this paper we present the results from our analysis of $\sim 53$
deg$^2$ of $R_C$-band data from the Red-Sequence Cluster Sequence
(RCS) (e.g., Yee \& Gladders 2001), which is a 90 deg$^2$ galaxy
cluster survey designed to provide a large sample of optically
selected clusters of galaxies with redshifts $0.1<z<1.4$.  

Hoekstra et al. (2002a) used a subset of the current data to study the
cosmic shear.  Compared to other cosmic shear studies, the RCS data
are shallow, and consequently the signal at a given scale is much
lower, as is the signal-to-noise ratio. However, measuring the weak
lensing signal from a shallow survey has several advantages. Down to a
limiting magnitude of $R_C\sim 24$ star-galaxy separation works well
(see Fig.~1 from Hoekstra et al 2002a). In deeper surveys many sources
have sizes comparable to the size of the PSF, and applying size cuts
may change the redshift distribution of the sources in a systematic
way. In addition, down to $R_C\sim 24$ the redshift distribution of
the sources is fairly well determined. In order to relate the observed
cosmic shear signal to cosmological parameters, a good understanding
of the source redshift distribution is crucial.

One worry is the effect of intrinsic alignments of the source
galaxies, which introduces an additional signal (e.g., Heavens et
al. 2000; Catelan et al. 2001; Crittenden et al. 2001; Mackey et
al. 2001). The amplitude of the effect is not well determined, but it
is clear that it is more important for shallower surveys. 
However, the predictions indicate that for a median redshift
of $z=0.6$ (which is similar to our sample of source galaxies) the
signal caused by intrinsic alignments is still small compared to the
lensing signal.

The structure of the paper is as follows. In \S2 we briefly
discuss the theory of lensing by large scale structure. The
data and measurements are presented in \S3. We compare
the measurements to model predictions in a maximum likelihood
analysis in \S4. We summarize our results in \S5.

\section{Method}

In this section we provide a basic description of the theory of weak
lensing by large scale structure. We discuss the dependence on the
assumed cosmology and the redshifts of the sources. Detailed
discussions on this subject can be found elsewhere (e.g., Schneider et
al. 1998; Bartelmann \& Schneider 2001).

The observable two-point statistics can be related to the 
convergence power spectrum, which is defined as 

\begin{equation} 
P_\kappa(l)=\frac{9 H_0^4 \Omega_m^2}{4 c^4}
\int\limits_0^{w_H}dw \left(\frac{\bar W(w)}{a(w)}\right)^2
P_\delta\left(\frac{l}{f_K(w)};w\right),
\end{equation}

\noindent where $w$ is the radial (comoving) coordinate, $w_H$
corresponds to the horizon, $a(w)$ the cosmic scale factor, and
$f_K(w)$ the comoving angular diameter distance. As shown by Jain \&
Seljak (1997) and Schneider et al. (1998) it is necessary to use the
non-linear power spectrum in equation~(1). This power spectrum is
derived from the linear power spectrum following the prescriptions
from Peacock \& Dodds (1996).

$\bar W(w)$ is the source-averaged ratio of angular diameter distances
$D_{ls}/D_{s}$ for a redshift distribution of sources $p_b(w)$:

\begin{equation}
\bar W(w)=\int_w^{w_H} dw' p_b(w')\frac{f_K(w'-w)}{f_K(w')}.
\end{equation}

Hence, it is important to know the redshift distribution of the
sources, in order to relate the observed lensing signal to
$P_\kappa(l)$. 

In Hoekstra et al. (2002a) we used the top-hat smoothed variance as a
measure of the cosmic shear signal. The top-hat variance
$\langle\gamma^2\rangle$ is related to the convergence power spectrum
through

\begin{equation}
\langle\gamma^2\rangle(\theta)=2\pi\int_0^\infty
dl~l~P_\kappa(l)\left[\frac{J_1(l\theta)}{\pi l \theta}\right]^2,
\end{equation}

\noindent where $\theta$ is the radius of the aperture used to compute
the variance, and $J_1$ is the first Bessel function of the first
kind. Hoekstra et al. (2002a) measured the top-hat variance by
actually computing the excess variance in apertures. 

Recent studies, however, show that an optimal use of the data is to
measure the shear correlation functions from the data directly. These
correlation functions can be related to the various two-point
statistics. In addition, this approach allows one to split the signal
into two components: an ``E''-mode, which is curl-free, and a
``B''-mode, which is sensitive to the curl of the shear field.
Gravitational lensing arises from a gravitational potential, and hence
it is expected to produce a curl-free shear field. Hence, the
``B''-mode can be used to quantify the level of systematics involved
in the measurement.

Several sources of ``B''-mode have been identified. Schneider et
al. (2002) showed how redshift clustering of source galaxies can
introduce a ``B''-mode. However, in most situations the contribution to
the observed signal is negligible. Simple models describing the
intrinsic alignments of galaxies predict a small ``B''-mode (Crittenden et
al. 2002), although the amplitude is still uncertain.  Hence, any
measured ``B''-mode is dominated by residual systematics in the data
(e.g., imperfect correction of the PSF anisotropy) or intrinsic
alignments. The effect of intrinsic alignments can be minimized by
selecting galaxies with a broad redshift distribution. In future
surveys, with photometric redshift information for the galaxies, the
contribution from intrinsic alignments can be removed completely by
correlating the shapes of galaxies with different redshifts.

The decompositions of the shear correlation function and the
top-hat variance in E and B modes are defined up to a constant
(Crittenden et al. 2002; Pen et al. 2002). The decomposition
is naturally carried out by using the aperture mass statistic
$M_{\rm ap}$. Kaiser et al. (1994) introduced this statistic
to measure the masses of clusters of galaxies. Its usefulness
for the study of cosmic shear was pointed out by Schneider et al.
(1998). The aperture mass is defined as 

\begin{equation}
M_{\rm ap}(\theta)=\int d^2\phi U(\phi) \kappa(\phi).
\end{equation}

\noindent Provided $U(\phi)$ is a compensated filter, i.e., $\int
d\phi \phi U(\phi)=0$, with $U(\phi)=0$ for $\phi>\theta$, the
aperture mass can be expressed in term of the observable tangential
shear $\gamma_{\rm t}$ using a different filter function $Q(\phi)$
(which is a function of $U(\phi)$),

\begin{equation}
M_{\rm ap}(\theta)=\int_0^\theta d^2\phi Q(\phi)\gamma_{\rm t}(\phi).
\end{equation}

We use the filter function suggested by Schneider et al. (1998)

\begin{equation}
U(\theta)=\frac{9}{\pi\theta^2}
\left(1-\frac{\vt^2}{\theta^2}\right)
\left(\frac{1}{3}-\frac{\vt^2}{\theta^2}\right),
\end{equation}

\noindent for $\theta\le\vt$, and 0 elsewhere. The corresponding $Q(\theta)$
is given by

\begin{equation}
Q(\theta)=\frac{6}{\pi\theta^2}\left(\frac{\vt^2}{\theta^2}\right)
\left(1-\frac{\vt^2}{\theta^2}\right),
\end{equation}

\noindent for $\theta\le\vt$, and 0 elsewhere. With this choice of
filter functions, the variance of the aperture mass 
$\langle M_{\rm ap}^2\rangle$ is related to the power spectrum through

\begin{equation}
\langle M_{\rm ap}^2\rangle=2\pi\int_0^\infty dl~l
P_\kappa(l)\left[\frac{12}{\pi (l \theta)^2} J_4(l \theta)\right]^2,
\end{equation}

\noindent where $J_4$ is the fourth-order Bessel function of the first
kind.

A straightforward implementation of the aperture mass statistic is
to compute $M_{\rm ap}$ directly from the data using the estimator
(e.g., Schneider et al. 1998)

\begin{equation}
\tilde M_{\rm ap}=\pi \theta^2
\frac{\sum_{i=1}^{N_s} Q(\theta_i) w_i \gamma_{{\rm t},i}}
{\sum_{i=1}^{N_s} w_i},
\end{equation}

\noindent where $N_s$ is the number of source galaxies, $\gamma_{{\rm t},i}$
is the tangential shear of the $i$th galaxy with respect to the
center of the aperture. The weights $w_i$ correspond to the inverse
square of the uncertainty in the shape measurement (Hoekstra et al. 2000).

One can use this procedure to tile the observed fields with apertures, and
compute the excess variance. However, this approach assumes that the
data are contiguous, which is not the case for real data.
The masking of the RCS data is not severe, but does complicate a measurement
of $\langle M_{\rm ap}^2\rangle$ through this method.

Instead, we will determine $\langle M_{\rm ap}^2\rangle$ from the
observed ellipticity correlation functions. This method has the
advantage that it uses all information contained in the data, and
that it does not depend on the geometry of the survey. Pen et al.
(2002) were the first to apply this approach, using data from the
VIRMOS-DESCART survey.

The two ellipticity correlation function that are measured are

\begin{equation}
\xi_{\rm tt}(\theta)=\frac{\sum_{i,j}^{N_s} w_i w_j 
\gamma_{{\rm t},i}({{\bf x}_i}) \cdot \gamma_{{\rm t},j}({{\bf x}_j})}
{\sum_{i,j}^{N_s} w_i w_j},
\end{equation}

\noindent and

\begin{equation}
\xi_{\rm rr}(\theta)=\frac{\sum_{i,j}^{N_s} w_i w_j 
\gamma_{{\rm r},i}({{\bf x}_i}) \cdot \gamma_{{\rm r},j}({{\bf x}_j})}
{\sum_{i,j}^{N_s} w_i w_j},
\end{equation}

\noindent where $\theta=|{\bf x}_i-{\bf x}_j|$. $\gamma_{\rm t}$ and
$\gamma_{\rm r}$ are the tangential and 45 degree rotated shear in the
frame defined by the line connecting the pair of galaxies. For the
following, it is more useful to consider

\begin{equation}
\xi_+(\theta)=\xi_{\rm tt}(\theta)+\xi_{\rm rr}(\theta),{\rm~and~}
\xi_-(\theta)=\xi_{\rm tt}(\theta)-\xi_{\rm rr}(\theta),
\end{equation}

\noindent i.e., the sum and the difference of the two observed
correlation functions. As shown by Crittenden et al. (2002), one can
derive ``E'' and ``B''-mode correlation functions by integrating
$\xi_+(\theta)$ and $\xi_-(\theta)$ with an appropriate window
function (see Pen et al. 2002 for an application to the VIRMOS-DESCART
data).

The ``E'' and ``B''-mode aperture masses are computed from the ellipticity correlation
functions using

\begin{equation}
\langle M_{\rm ap}^2\rangle(\theta)=\int d\vt~\vt \left[{\cal W}(\vt)\xi_+(\vt)+
\tilde{\cal W}(\vt)\xi_-(\vt)\right],
\end{equation}

\noindent and

\begin{equation}
\langle M_\perp^2\rangle(\theta)=\int d\vt~\vt \left[{\cal W}(\vt)\xi_+(\vt)-
\tilde{\cal W}(\vt)\xi_-(\vt)\right],
\end{equation}

\noindent where ${\cal W}(\vt)$, and $\tilde{\cal W}(\vt)$ are given
in Crittenden et al. (2002). Useful analytic expressions were derived
by Schneider et al. (2001). Both ${\cal W}(\vt)$, and $\tilde{\cal
W}(\vt)$ vanish for $\vt>2\theta$, so that $\langle M_{\rm
ap}^2\rangle$ can be obtained directly from the observable ellipticity
correlation functions over a finite interval.

\begin{figure*}[t!]
\begin{center}
\leavevmode 
\hbox{%
\epsfxsize=0.85\hsize 
\epsffile[18 315 558 710]{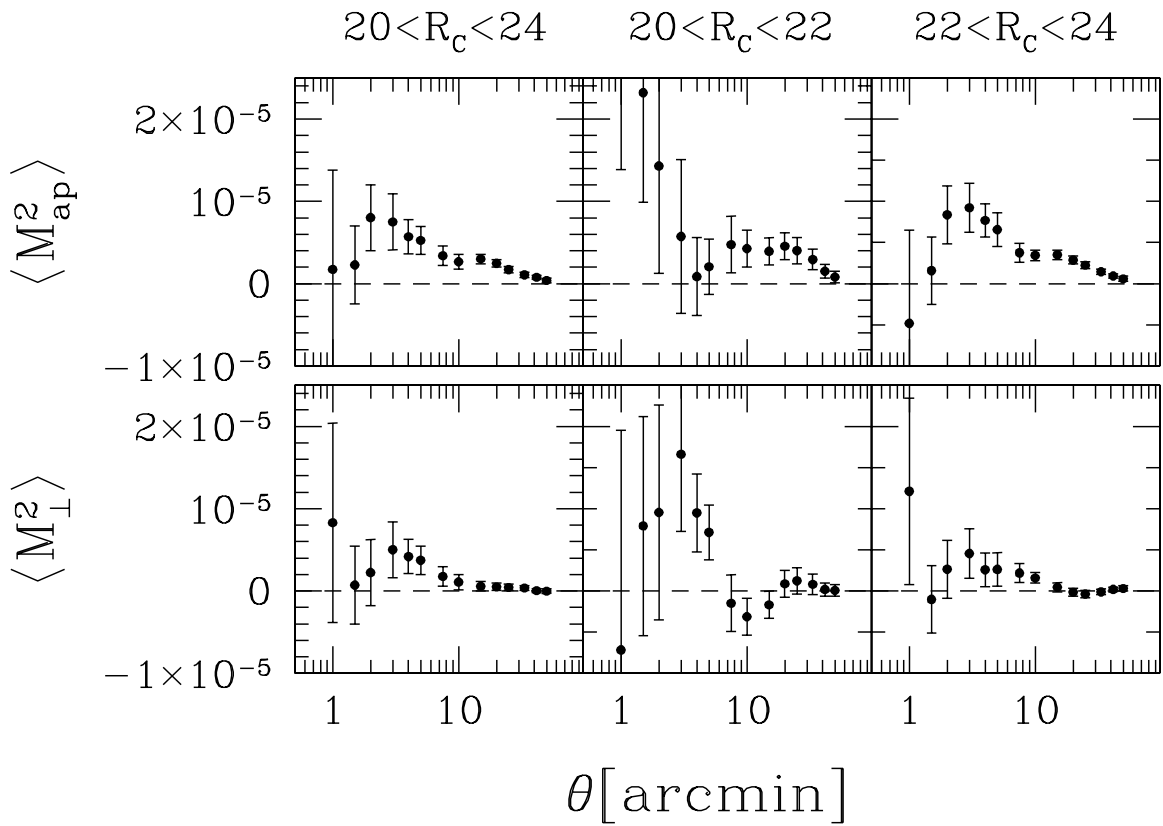}} 
\figcaption{\footnotesize The upper panels show the measured variance
of the aperture mass $\langle M_{\rm ap}^2\rangle$ as a function of
aperture size $\theta$ for different samples of source galaxies.  This
signal corresponds to the ``E'-mode. The lower panels show the variance
$\langle M_\perp^2\rangle$ when the phase of the shear is increased by
$\pi/2$, and corresponds to the ``B''-mode. The error bars indicate the
$1\sigma$ statistical uncertainty in the measurements, and have been
derived from the field-to-field variation of the 13 patches (thus the
error bars include cosmic variance). Note that the points are slightly
correlated. We detect a significant ``B''-mode on scales $5-10$
arcminutes. On scales larger than 10 arcminutes the ``B''-mode
vanishes. The sample of bright galaxies $(20<R_C<22)$ should not be
affected significantly by systematics because their sizes are large
compared to the PSF. Therefore the significant ``B''-mode at scales of a
few arcminutes is likely to be caused by intrinsic alingments.  To
minimize the effect of intrinsic alignments on our cosmological
parameter estimation, we will use the sample of galaxies with
$22<R_C<24$ to this end.
\label{map}}
\end{center}
\end{figure*}

\section{Measurements}

We use the $R_C$-band imaging data from the Red-Sequence Cluster
Survey (RCS, Gladders \& Yee 2000). The final survey covers 90 deg$^2$
in both $R_C$ and $z'$, spread over 22 widely separated patches of
$\sim 2.1\times 2.3$ degrees. The northern half of the survey was
observed using the CFH12k camera on the CFHT, and the data from the
southern half were obtained using the Mosaic II camera on the CTIO 4m
telescope.

The integration times are 900s for the CFHT data and 1200s for the
CTIO data. A subset of this data set was studied in Hoekstra et
al. (2002a): 16.4 deg$^2$ CFHT data and 7.6 deg$^2$ of CTIO data. We
augment the data used by Hoekstra et al. (2002a) by the remaining
CFHT12k data. Much of the remaining CTIO data has relatively poor
seeing, and hence is less useful for our weak lensing analysis and is
not used. The seeing for the data used here ranges from $0\farcs{57}$
to $1\farcs{39}$ (only 11\% of the data had seeing $>1''$), with a
median value of $0\farcs{8}$. Hence, the final data set studied here
consists of a total of 53 deg$^2$ of imaging data, spread over 13
patches.

A detailed description of the data reduction and object analysis
is described in Hoekstra et al. (2002a), to which we refer for
technical details. Here we present a short description of the
various steps in the analysis.

We use single exposures in our analysis, and consequently cosmic rays
have not been removed. However, cosmic rays are readily eliminated
from the photometric catalogs: small, but very significant objects are
likely to be cosmic rays, or artifact from the CCD. The peak finder
gives fair estimates of the object size, and we remove all objects
smaller than the size of the PSF.

The objects in this cleaned catalog are then analysed, which yield
estimates for the size, apparent magnitude, and shape parameters
(polarization and polarizabilities). The objects in this catalog are
inspected by eye, in order to remove spurious detections.  These
objects have to be removed because their shape measurements are
affected by cosmetic defects (such as dead columns, bleeding stars,
halos, diffraction spikes) or because the objects are likely to be
part of a resolved galaxy (e.g., HII regions). 

To measure the small, lensing induced distortions in the images of the
faint galaxies it is important to accurately correct the shapes for
observational effects, such as PSF anisotropy, seeing and camera
shear; PSF anisotropy can mimic a cosmic shear signal, and a
correction for the seeing is required to relate the measured shapes to
the real lensing signal. To do so, we follow the procedure outlined in
Hoekstra et al. (1998).  We select a sample of moderately bright stars
from our observations, and use these to characterize the PSF
anisotropy and seeing.  We fit a second order polynomial to the shape
parameters of the selected stars for each chip. These results are used
to correct the shapes of the galaxies for PSF anisotropy and seeing.

The effect of the PSF is not the only observational distortion that
has to be corrected. The optics of the camera stretches the images of
galaxies (i.e., it introduces a shear) because of the non-linear
remapping of the sky onto the CCD.  We have used observations of
astrometric fields to find the mapping between the sky and the CCD
pixel coordinate system, and derived the corresponding camera shear,
which is subsequently subtracted from the galaxy ellipticity (see
Hoekstra et al. 1998).

As discussed in Hoekstra et al. (2002a) we use galaxies with
$20<R_C<24$ in our analysis. This selection yields a sample
of 1,773,543 galaxies for which we could determine useful shape parameters.
We note that the sample is not complete at the faint limit.

The upper panels in Figure~\ref{map} show the measured variance of the
aperture mass $\langle M_{\rm ap}^2\rangle$ as a function of aperture
size $\theta$. For all three samples we find that the ``B''-mode
(lower panels) vanishes on scales larger than $\sim 10$ arcminutes,
suggesting that neither observational distortions or intrinsic
alignments of sources have corrupted our measurements. This is quite
different from the VIRMOS-DESCART results presented by Van Waerbeke et
al. (2002) who found an almost constant ``B''-mode out to scales of 30
arcminutes.

To study possible residual systematics introduced by the correction
for the PSF and camera shear, we split the data into various subsets:
good/bad seeing, small/large PSF anisotropy, center/edge of chip.
In all three cases, we find excellent agreement between the subsets.

On smaller scales we detect a significant ``B'' mode. Interestingly,
we detect also a significant signal for the bright galaxies
$(20<R_C<22)$. These galaxies have sizes that are large compared to
the PSF, and therefore they are less affected by residual systematics,
whereas intrinsic alignments are expected to be particularly important
for bright galaxies. The selection of relatively bright galaxies in
the $R_C$ band results in a relatively narrow range in redshift as one
essentially observes intrinsically fainter galaxies at redshifts
$z=0.3-0.5$. Only for fainter magnitude limits one overcomes the
4000\AA~break, and more distant galaxies enter in the sample. Hence,
the ``B''-mode at a few arcminutes is likely to be caused by intrinsic
alignments.

\vbox{
\begin{center}
\leavevmode 
\hbox{%
\epsfxsize=8.5cm 
\epsffile{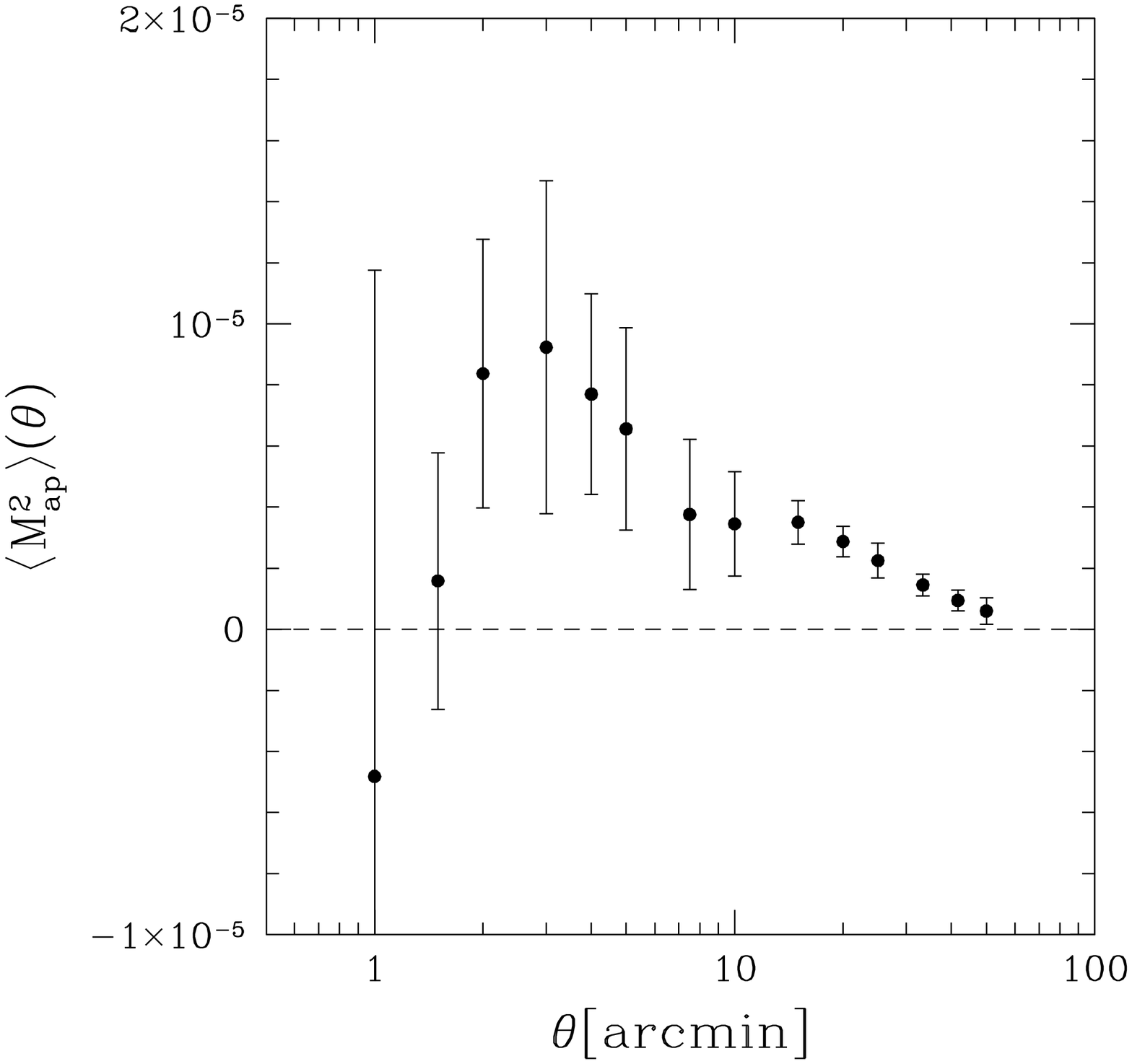}}
\figcaption{\footnotesize The measured variance
of the aperture mass $\langle M_{\rm ap}^2\rangle$ as a function of
aperture size $\theta$ for source galaxies with $22<R_C<24$. The
error bars have been increased to account for the unknown correction
for the ``B''-mode observed in Figure~\ref{map}. 
\label{mapfaint}}
\end{center}}

The full sample $(20<R_C<24)$ includes the bright galaxies, and hence
the contribution of intrinsic alignments to the signal is larger
compared to the faint $(22<R_C<24$) sample. The fainter galaxies yield
a slightly higher lensing signal, because the average redshift of the
sources is higher. In addition the redshift distribution of galaxies
with $22<R_C<24$ is also broader (see, e.g., Hoekstra 2001), which
reduces the effect of intrinsic alignments.

Currently it is not clear how to correct the ``E''-mode, given the
observed ``B''-mode. We follow Van Waerbeke et al (2002) and add
the ``B''-mode in quadrature to the uncertainty in the ``E''-mode,
as this provides a conservative limit to how well we can determine
the lensing signal. The best signal-to-noise is obtained for the
faint sample, and we will use this sample below for the cosmological
parameter estimation.

Figure~\ref{mapfaint} shows the resulting measurement of  
$\langle M_{\rm ap}^2\rangle$ as a function of scale when the
error bars are increased using the observed ``B''-mode. 

\section{Constraints on cosmological parameters}

\subsection{Source redshift distribution}

We now need to relate the observed lensing signal, presented in
Figure~\ref{mapfaint}, to the cosmological parameters. To do so, we
compute model predictions for a range of parameters using
Eq.~(8). However, the evaluation of Eq.~(8) requires knowledge of the
redshift distribution of the sources.

Compared to other, deeper, cosmic shear studies, the RCS data have the
major advantage that the redshift distribution of the galaxies is
known fairly well.  Cohen et al. (2000) have determined
spectroscopically the redshift distribution of galaxies down to a
limiting magnitude of $R_C=24$. However, their result is likely to
suffer from sample variance. Comparison of the Cohen et al. (2000)
redshifts with photometric redshifts (e.g., Fern{\'a}ndez-Soto et
al. 1999) showed that the latter provide a suitable way to derive the
redshift distribution of the sources we consider here.

To do so, we have to account for the fact that the uncertainty in the
shape measurements depend on the apparent magnitudes (and thus on the
redshifts) of the sources: the contribution of distant, small faint
galaxies (with noisy shape measurements) to the measured lensing
signal is smaller compared to brighter galaxies. The relative weight
as a function of magnitude (see Figure~7 in Hoekstra et al. 2002a), is
used to derive the ``effective'' redshift distribution. It turns out
that this weighting scheme changes the redshift distribution only
slightly. We refer to Hoekstra et al. (2000; 2002a) for a more
detailed discussion of the photometric redshift distribution used here
(also see Van Waerbeke et al. 2002).

It is expected that our knowledge of the redshift distribution
of galaxies brighter than $R_C=24$ will improve significantly
in the near future. The current distribution is based on a
relatively small sample of galaxies, and hence the data allow
for a (small) range in redshift distributions.

To allow for this uncertainty in the redshift distribution we marginalize
our model predictions over the redshift distribution allowed by the
measurements of Fern{\'a}ndez-Soto et al. (1999). To do so, we 
parametrize the redshift distribution using

\begin{equation}
p_b(z)=\frac{\beta}{z_s \Gamma\left(\frac{1+\alpha}{\beta}\right)}
\left(\frac{z}{z_s}\right)^\alpha \exp\left[-\left(\frac{z}{z_s}\right)^\beta
\right].
\end{equation}

\begin{figure*}[t!]
\begin{center}
\leavevmode 
\hbox{%
\epsfxsize=6cm 
\epsffile{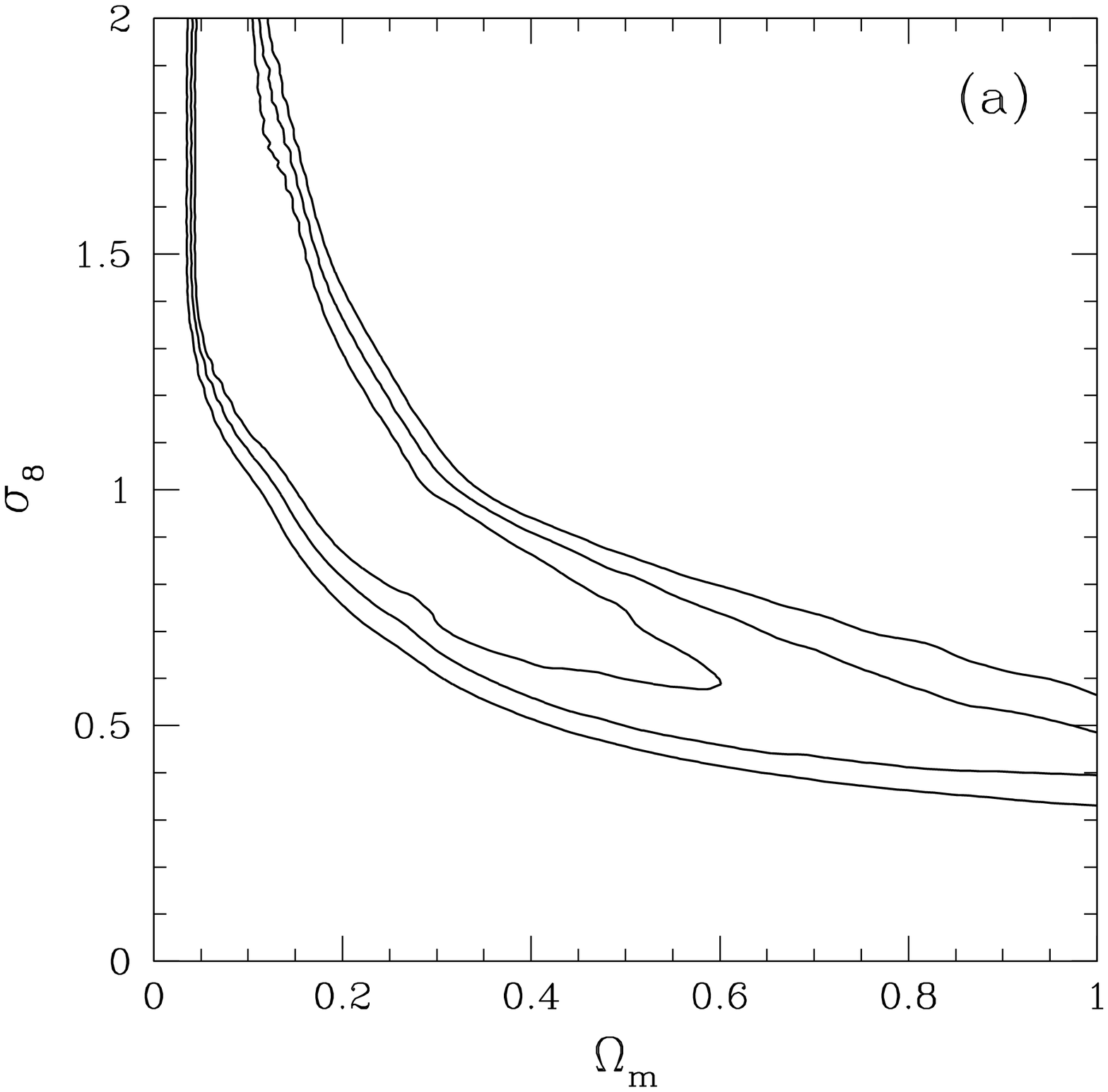}
\epsfxsize=6cm 
\epsffile{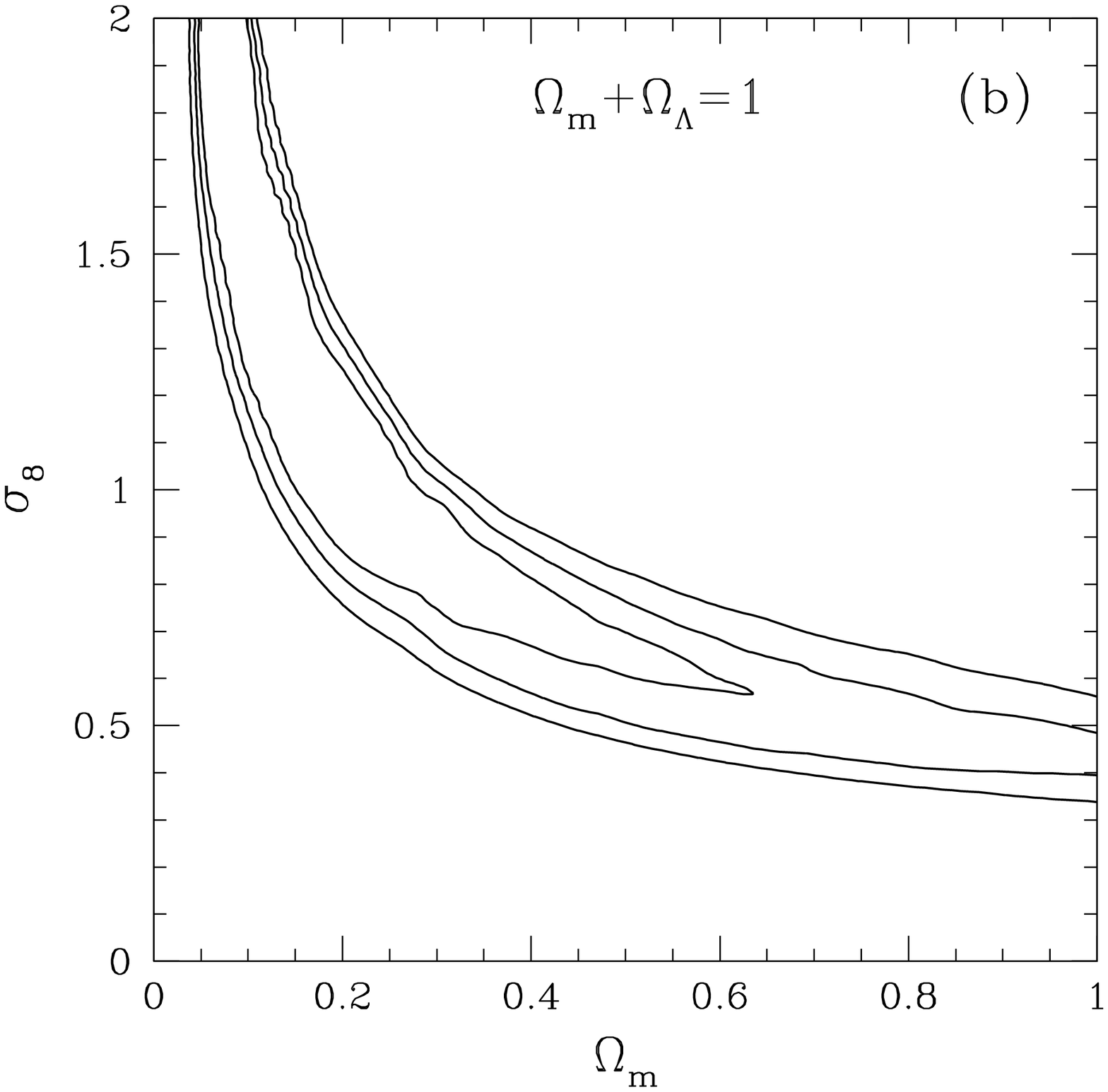}
\epsfxsize=6cm 
\epsffile{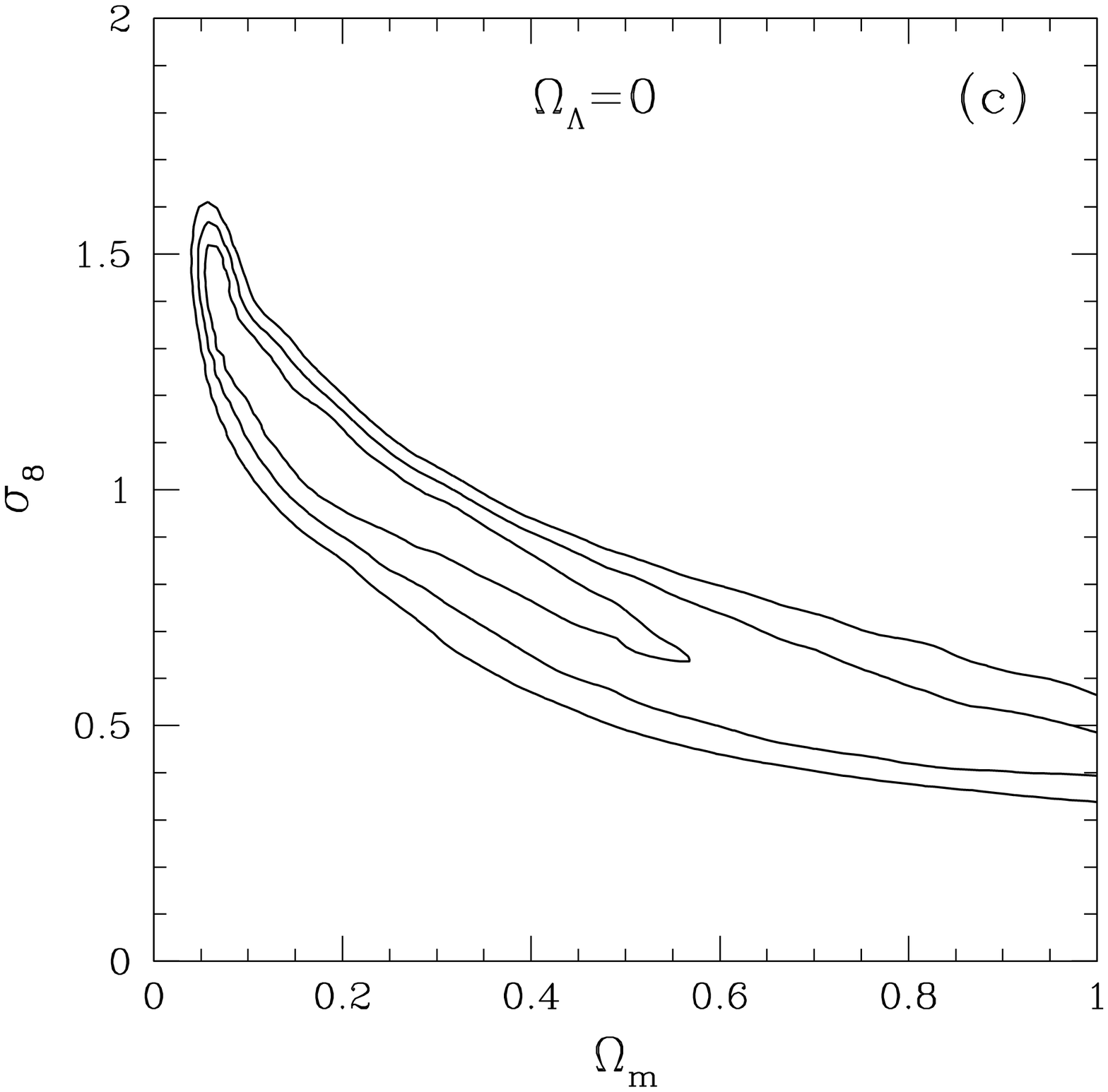}}
\figcaption{\footnotesize Contraints on $\Omega_m$ and $\sigma_8$ from
RCS data only. The likelihood contours have been derived by comparing
the measurements to CDM models with $n=1$. We have marginalized the
models over $\Gamma$ and the $z_s$ (for which we assumed
a flat prior). The contours indicate the 68.3\%, 95.4\%, and 99.7\%
confidence limits on two paramaters jointly. Panel~(a) shows the
results marginalized over $\Omega_\Lambda$; (b) gives the constraints
for a flat cosmology (as suggested by the recent CMB results); (c)
the results for an open  $(\Omega_\Lambda=0)$ cosmology.
\label{cont_om_s8}}
\end{center}
\vspace{-0.5cm}
\end{figure*}

We fit this distribution to the observed ``effective'' redshift
distribution, fixing $\alpha=4.7$ and $\beta=1.7$, which yields a best
fit value of $z_s=0.302$.  We compute models for
$z_s\in[0.274,0.337]$, which correspond to the $\pm 3\sigma$ range
indicated by $\chi^2$ of the fit to the photometric redshift
distribution. For this choice of parameters the range in mean source
redshift is $\langle z\rangle=0.54-0.66$.

\subsection{Parameter estimation}

In this section we briefly discuss how we obtain constraints on some
of the cosmological parameters. The relevant parameters for the
analysis presented here are the mass density $\Omega_m$, the
normalization of the power spectrum $\sigma_8$ and the shape parameter
$\Gamma$ (note that $\Gamma\sim\Omega_m h$ in CDM cosmologies). 

We also consider $\Omega_\Lambda$, although the RCS results are not
expected to provide strong constraints on $\Omega_\Lambda$, because
lensing is not sensitive to this parameter. However, the combination
of the RCS measurements with the VIRMOS-DESCART results (Van Waerbeke
et al. 2001a; 2002) can provide stronger constraints, because the two
surveys probe the matter power spectrum at different redshifts. Hence,
such a combined analysis enables a measurement of the growth of
structure, and as a result it can break some of the degeneracies
limiting our analysis.  Furthermore, we can impose priors on some of
the parameters based on external measurements, such as the CMB or
redshift surveys.

We compute model predictions on a grid, which are subsequently
compared to the data. We have limited the model parameters to
a realistic range. We consider $\Omega\in [0,1]$, $\Omega_\Lambda\in [0,1]$,
$\sigma_8\in [0,2]$, and $\Gamma\in [0.05,0.5]$. As mentioned above,
we use $z_s\in [0.274,0.337]$ to account for the uncertainty in the
source redshift distribution.

The measurements of $\langle M_{\rm ap}^2\rangle$ at different scales
are slightly correlated. In order to compare the measurements to the
models we have to compute the covariance matrix ${\bf C}$ which is
given by

\begin{equation}
C_{ij}=\langle (d_i-\mu_i)^T(d_j-\mu_j)\rangle,
\end{equation}

\noindent where $d_i$ is the measurement at scale $\theta_i$, and
$\mu_i$ is the ``true'' value for this scale (for which we take
ensemble averaged value from the data). We compute the covariance
matrix from the variance in the measurements for the 13 different
patches, and hence it includes the cosmic variance. For a given model,
we can compute the $\chi^2$ (log-likelihood)

\begin{equation}
\chi^2=(d_i-m_i){\bf C}^{-1}(d_i-m_i)^T,
\end{equation}

\noindent which can be used to derive the likelihood for each set of
parameters.  From the measured $\chi^2$ values it is straightforward
to calculate the confidence contours.

\subsection{Joint constraints on $\Omega_m$ and $\sigma_8$}

We first consider the constraints we can derive on the combination
of $\Omega_m$ and $\sigma_8$ from the RCS data alone. Without
external priors these two cosmological parameters are degenerate.
However, studies of cluster abundances give similar constraints, and
it is therefore useful to compare the lensing results to the latter,
popular technique.

Figure~\ref{cont_om_s8} shows the joint constraints on $\Omega_m$ and
$\sigma_8$. We have marginalized over $\Gamma$ and $z_s$, for which we
assumed flat priors (i.e., all values have equal likelihood), which is
rather conservative.  Panel~a shows the most general result, where we
marginalize over $\Omega_\Lambda$ as well. For this choice of priors,
we obtain $\sigma_8=0.45^{+0.09}_{-0.12}\Omega_m^{-0.55}$ (95\%
confidence).

The current CMB measurements favor a flat cosmology (e.g., de
Bernardis et al. 2000; Jaffe et al. 2001; Netterfield et al. 2001;
Pryke et al. 2002; Stompor et al. 2001), and the results for these
models are presented in panel~b. For reference we also show the
results for an open $(\Omega_\Lambda=0)$ universe.

The current constraints on $\Omega_m$ and $\sigma_8$ in
Figure~\ref{cont_om_s8} are degenerate from lensing alone. The
degeneracies can be broken by measuring the cosmic shear signal in
both the linear and the non-linear regime (Jain \& Seljak,
1997). However, the accuracy of our measurements at small scales
(non-linear regime) is limited by the observed ``B''-mode in the
data. Hence, due to the increased uncertainties on small scales,
we cannot break the degeneracy. 

However, if the ``B''-mode is caused by intrinsic alignments of
galaxies, forthcoming multi-color data can be used to derive crude
photometric redshifts, and we might be able to remove the contribution
of intrinsic alignments by correlating the shapes of galaxies at
different redshifts. In addition, the planned CFHT Legacy Survey,
which is a 170 deg$^2$ deep, multi-color survey will allow us to
measure the cosmic shear signal with unprecedented accuracy, and
these data will allow a determination of $\Gamma$, $\sigma_8$,
$\Omega_m$ and $\Omega_\Lambda$ from the shear two-point functions
alone.

We also note that in order to break the degeneracies, and infer
the correct values for the various cosmological parameters,
the prescription for the evolution of the non-linear power spectrum
needs to be improved. Van Waerbeke et al. (2001b, 2002) showed
that the non-linear predictions fail for some cosmologies. The
current prescription (Peacock \& Dodds 1996) is not accurate
enough for the future measurements.

Fortunately, cosmological parameters can be estimated through a wide range
of techniques, such as the fluctuations in the CMB and galaxy redshift
surveys. We can use the most recent results as priors to obtain
tighter constraints.

In Figure~\ref{priors} we show the constraints on $\Omega_m$ and
$\sigma_8$, when we use Gaussian priors for the source redshift
distribution $z_s$, $\Gamma$ as determined from the 2dF galaxy
redshift survey (Peacock et al. 2001; Efstathiou et al. 2001), and 
$\Omega_{\rm tot}=\Omega_m+\Omega_\Lambda$ from the Boomerang CMB
measurements (Netterfield et al. 2001).  The contours have tightened
considerably, and the data favor a low value for $\Omega_m$. For this
choice of priors, we find $\sigma_8=(0.46^{+0.05}_{-0.07})\Omega_m^{-0.52}$
(95\% confidence).

We have also investigated whether the RCS data can be used to
constrain $\Gamma$. Using our set of Gaussian priors, we find
that we can only place a lower bound on $\Gamma$. We find
a 95\% confidence lower limit of $\Gamma>0.1+0.16\Omega_m$.

\vbox{
\begin{center}
\leavevmode 
\hbox{%
\epsfxsize=8.0cm 
\epsffile{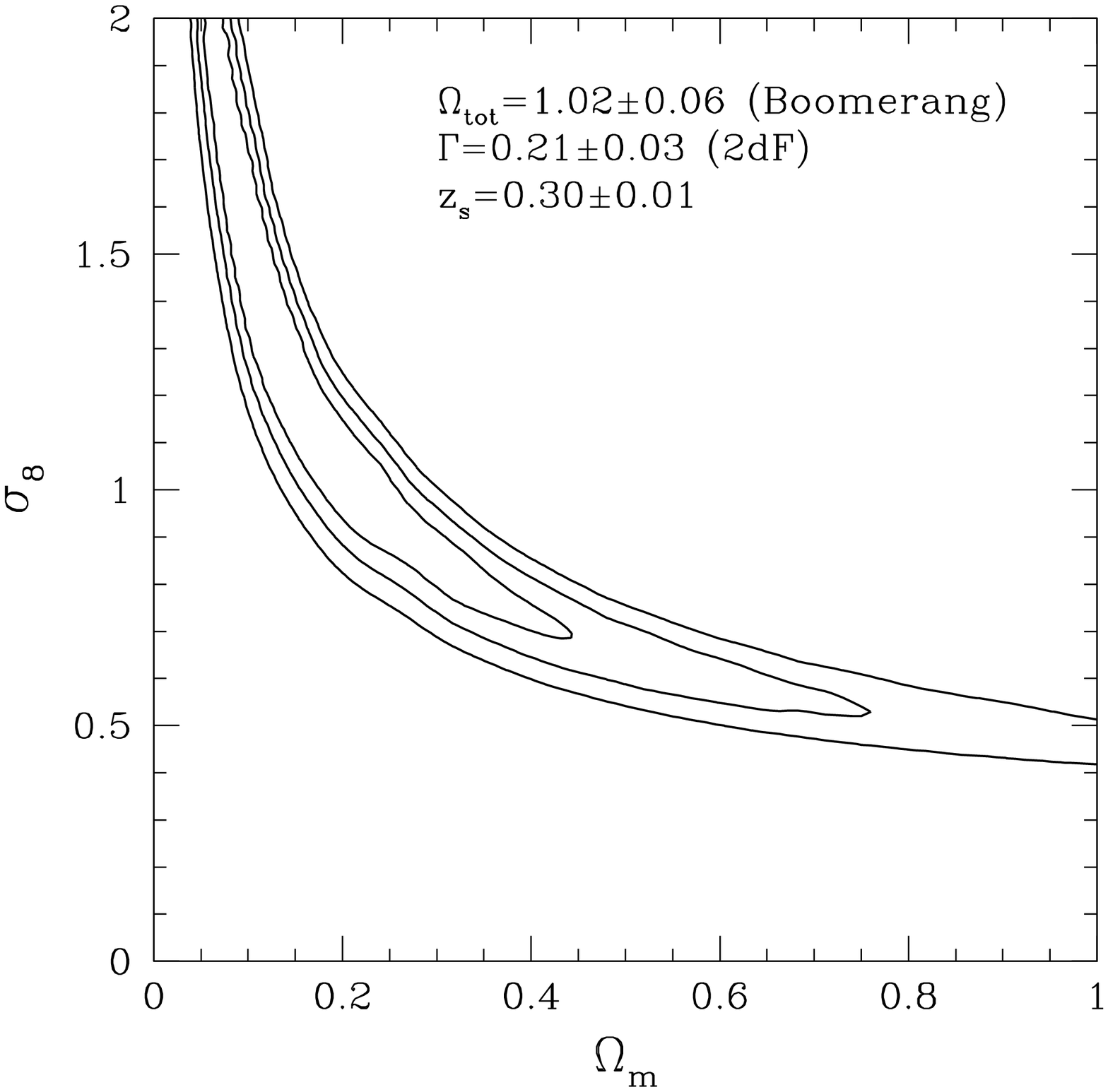}}
\figcaption{\footnotesize 
Contraints on $\Omega_m$ and $\sigma_8$ using Gaussian priors
on $\Gamma$, $\Omega_{\rm tot}=\Omega_m+\Omega_\Lambda$, and the source
redshift distribution $z_s$ (which gives $\langle
z\rangle=0.59\pm0.02$).  The likelihood contours have been derived by
comparing the measurements to CDM models with $n_s=1$.  For $\Gamma$ we
used the constraints from the 2dF survey $\Gamma=0.21\pm0.03$ (Peacock
et al. 2001; Efstathiou et al. 2001), and for $\Omega_{\rm tot}$ we used the
Boomerang constraints $\Omega_{\rm tot}=1.02\pm0.06$ (Netterfield et
al. 2001). The contours indicate the 68.3\%, 95.4\%, and 99.7\%
confidence limits on two paramaters jointly.
\label{priors}}
\end{center}}

\vbox{
\begin{center}
\begin{tabular}{ll}
\hline
\hline
$\sigma_8$	& priors \\
\hline
$0.88^{+0.04}_{-0.16}$ & flat prior on $z_s$ \\
$0.91^{+0.05}_{-0.12}$ & flat prior on $z_s$, $\Omega_m+\Omega_\Lambda=1$ \\
$0.94^{+0.04}_{-0.06}$ & flat prior on $z_s$, $\Omega_\Lambda=0$ \\
$0.86^{+0.04}_{-0.05}$ & Gaussian priors on $z_s$, $\Gamma$, and $\Omega_{\rm tot}$\\
\hline
\hline
\end{tabular}
\tabcaption{\footnotesize Values of $\sigma_8$ and 68\% confidence
intervals for different priors, assuming $\Omega_m=0.3$.
\label{tab_s8}}
\end{center}}

\vbox{
\begin{center}
\begin{tabular}{lll}
\hline
\hline
(a) & Cosmic shear survey & $\sigma_8$ \\
\hline
 & RCS				& $0.86^{+0.04}_{-0.05}$ \\
 & Bacon et al. (2002)		& $0.97^{+0.10}_{-0.09}$ \\
 & Refregier et al. (2002)	& $0.94\pm 0.14$\\
 & Van Waerbeke et al. (2002)	& $0.98\pm 0.06$ \\
\hline
\hline
(b) & Cluster abundance & $\sigma_8$ \\
\hline
 & Eke, Cole \& Frenk (1996)	& $0.93\pm0.07$\\
 & Carlberg et al. (1997)	& $0.82\pm0.03$\\
 & Bahcall \& Fan (1998)	& $1.18^{+0.24}_{-0.22}$\\
 & Pen (1998)			& $1.00\pm0.09$\\
 & Borgani et al. (2001)	& $0.72^{+0.07}_{-0.05}$\\
 & Pierpaoli et al. (2001)	& $1.02^{+0.07}_{-0.08}$ \\
 & Reiprich \& B{\"o}hringer (2001) & $0.68^{+0.08}_{-0.06}$ \\
 & Seljak et al. (2001)		& $0.75\pm0.06$ \\
 & Viana et al. (2001)		& $0.61\pm0.05$ \\
\hline
\hline
\end{tabular}
\tabcaption{\footnotesize (a) Values of $\sigma_8$ and 68\% confidence
intervals as derived from 4 independent cosmic shear measurements (adopting
$\Omega_m=0.3$, $\Omega_\Lambda=0.7$, and $\Gamma=0.21$; (b) 
determinations of $\sigma_8$ using cluster abundances.
\label{tab_comp}}
\end{center}}

For comparison to other work, we list the 68\% confidence limits on
$\sigma_8$ for $\Omega_m=0.3$ for the different priors from our work
in Table~\ref{tab_s8}. It is also interesting to compare our results
to other measurements. In Table~\ref{tab_comp}a we list our
measurements with the most recent measurements of $\sigma_8$ from weak
lensing by large scale structure. The agreement between the four
independent results is remarkable, in particular given the fact that
the measurements are based on a wide variety of data sets, using
different telescopes, filters, and depths.  We can compute the
ensemble averaged value of $\sigma_8$ from weak lensing, for which we
find a value of $0.92\pm0.03$.

We note that the value of $\sigma_8$ determined from the RCS data
is essentially constrained by the measurements at scales larger
than 10 arcminutes, where the ``B''-mode is negligible. The result
from Van Waerbeke et al. (2002) might be biased high, because
of their large scale ``B''-mode. Both Bacon et al. (2002) and Refregier
et al. (2002) do not separate their signal into ``E'' and ``B''-modes,
and therefore it might also include some residual systematics. 

A widely used method to determine the normalization of the power
spectrum uses the number density of rich clusters of galaxies (e.g.,
Borgani et al. 2001; Carlberg et al. 1997, Eke, Cole \& Frenk 1998;
Bahcall \& Fan 1998; Pen 1998; Pierpaoli, Scott, \& White 2001;
Reiprich \& B{\"o}hringer 2001; Seljak 2001; Viana, Nichol, \& Liddle
2001). Such systems are rare, and as a result a very sensitive probe
of $\sigma_8$, provided one can determine their
mass. Table~\ref{tab_s8}b lists the results from 9 of these
studies. The derived values for $\sigma_8$ from this technique range
from values as low as $0.61\pm0.05$ (Viana et al. 2001) to values
around unity (e.g., Bahcall \& Fan 1998; Pen 1998; Pierpoali 2001).
Unfortunately, it is currently unclear why some of the recent values
from cluster abundances are this low. From Table~\ref{tab_s8}b it is
clear that there is a wide spread in values, whereas the statistical
error bars are small. This points to underlying systematics, which
complicate the determination of $\sigma_8$ from cluster abundances.

\vbox{
\begin{center}
\leavevmode 
\hbox{%
\epsfxsize=8.5cm 
\epsffile{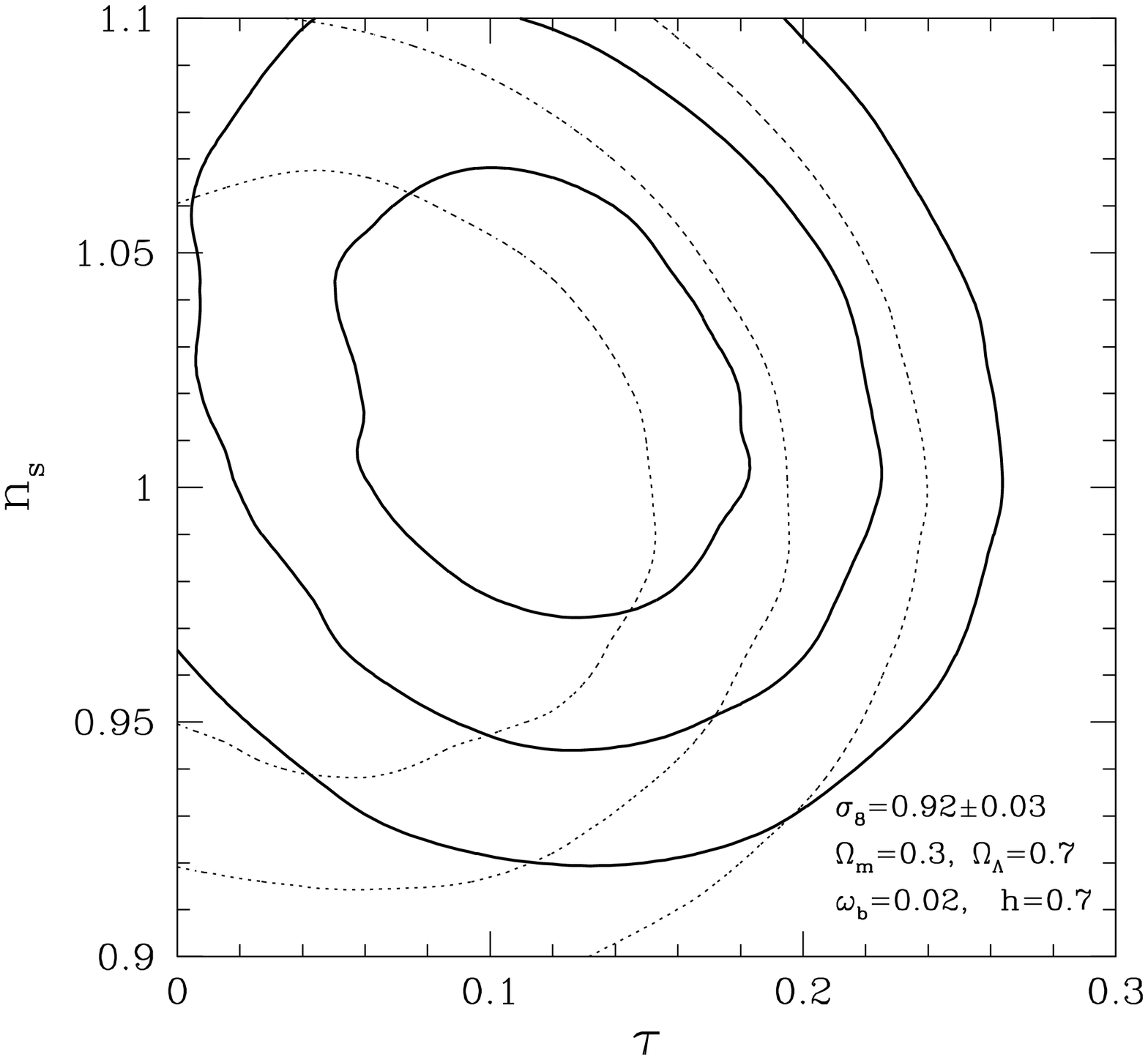}}
\figcaption{\footnotesize Joint constraints on the reionization
optical depth $\tau$ and the scalar spectral index $n_s$ from
the combination of CMB results (compilation by Wang et al. 2001).
We have marginalized over the value of $\sigma_8$, using a
prior $\sigma_8=0.92\pm0.03$ as determined by the 4 most recent
weak lensing measurements. The other cosmological parameters
were fixed as indicated. The dotted contours indicate the
constraints from RCS alone ($\sigma_8=0.86\pm0.05$).
The contours correspond to the  68.3\%, 95.4\% and 99.7\% confidence limits 
on the two parameters jointly.
\label{tau}}
\end{center}}

\subsection{Limits on the reionization optical depth $\tau$}

The measurements of the fluctuations in the CMB also give joint
constraints on $\Omega_m$ and $\sigma_8$. Recently, several papers
argue that the CMB results yield a low normalization for the matter
power spectrum (e.g., Lahav et al. 2001; Melchiorri \& Silk 2002),
with values ranging from $0.73-0.8$.

However, the current CMB data give relatively weak constraints on
$\sigma_8$, because the amplitude of the CMB power spectrum also
depends on the reionization optical depth $\tau$, and the scalar
spectral index $n_s$. Once the universe is reionized, CMB photons can
be Thomson scattered off of a free electron. As a result, the accousic
peaks in the CMB power spectrum are suppressed. Although the effect is
scale dependent (it does not affect the largest scales), it introduces
a degeneracy between $\sigma_8$ and $\tau$.  The matter power spectrum
probed by weak lensing is not affected, and hence, the combination
of the CMB and weak lensing results, in principle, can constrain
$\tau$, and consequently the reionization redshift.

The CMB results are taken from the compilation by Wang, Tegmark, \&
Zaldarriaga (2001), which is a combination of many experiments, such
as CBI (Padin et al. 2001), BOOMERANG (Netterfield et al. 2001), DASI
(Halverson et al. 2002), and MAXIMA (Lee et al. 2001). We use these
results to obtain joint constraints on $\tau$, $n_s$ and $\sigma_8$,
adopting a baryon density $\omega_b=0.02 $, $h=0.7$, $\Omega_m=0.3$,
and $\Omega_\Lambda=0.7$. We use the CMBFAST code (Seljak \&
Zaldarriaga 1996) to compute the theoretical CMB power spectra, and
use the window function and covariance matrix provided by Wang et
al. (2001).

In Figure~\ref{tau} we present the joint constraints on $\tau$ and
$n_s$, where we marginalized over the value of $\sigma_8$, using the
prior $\sigma_8=0.92\pm0.03$, as determined from weak lensing.  The
other cosmological parameters were fixed. The dotted contours
correspond to the constraints derived from the RCS measurement of
$\sigma_8$ alone. We stress that the results depend on the assumed
priors (many cosmological parameters are fixed), and a full combined
analysis of weak lensing and CMB data, marginalizing over all
parameters, is required for definite results. However, the results
presented here demonstrate that such a combined analysis can be
fruitful.

For our choice of parameters, which are close to those
favored by the vast amount of data currently available, we can derive
constraints on $n_s$ and $\tau$. Marginalizing over $\tau$, we find
$n_s=1.02^{+0.07}_{-0.06}$ with 95\% confidence. For $\tau$ we
obtain $\tau=0.12\pm0.09$ (95\% confidence, marginalized over $n_s$).

The reionization optical depth can be related to the redshift of
reionization. To do so, we use (e.g., Hu 1995; Griffiths \& Liddle
2001)

\begin{eqnarray}
\tau(z)= && 4.61\times 10^{-2}\left(1-\frac{Y_p}{2}\right) x_e
\frac{\Omega_b h}{\Omega_m}\times \nonumber\\
&& \left[\sqrt{1-\Omega_m+\Omega_m (1+z)^3}-1\right],
\end{eqnarray}

\noindent which is valid for flat cosmologies. Here $Y_p=0.24$ is the
primordial mass fraction in Helium, and $x_e$ is the ionized
fraction. For a fully reionized universe $(x_e=1)$ the resultant
$\tau$ as a function of reionization redshift is presented in
Figure~\ref{tau_z}. We have also indicated the 68\% and 95\%
confidence limits on $\tau$ (hatched regions), based on our
constraints on $\tau$. This allows us to read off the reionization
redshift $z_{\rm reion}$, for which we find $z_{\rm
reion}=14^{+7}_{-9}$ with 95\% confidence.

\section{Conclusions}

We have analysed $\sim 53$ square degrees of $R_C$-band imaging data
from the Red-Sequence Cluster Survey (RCS), and measured the excess
correlations in the shapes of galaxies on scales out to one degree. To
this end we use the aperture mass statistic, which allows a natural
separation of the signal in a curl-free ``E''-mode (which is expected
from gravitational lensing) and a ``B''-mode which provides an
important measure of residual systematics and the contribution from
intrinsic alignments.

\vbox{
\begin{center}
\leavevmode 
\hbox{%
\epsfxsize=8.5cm 
\epsffile{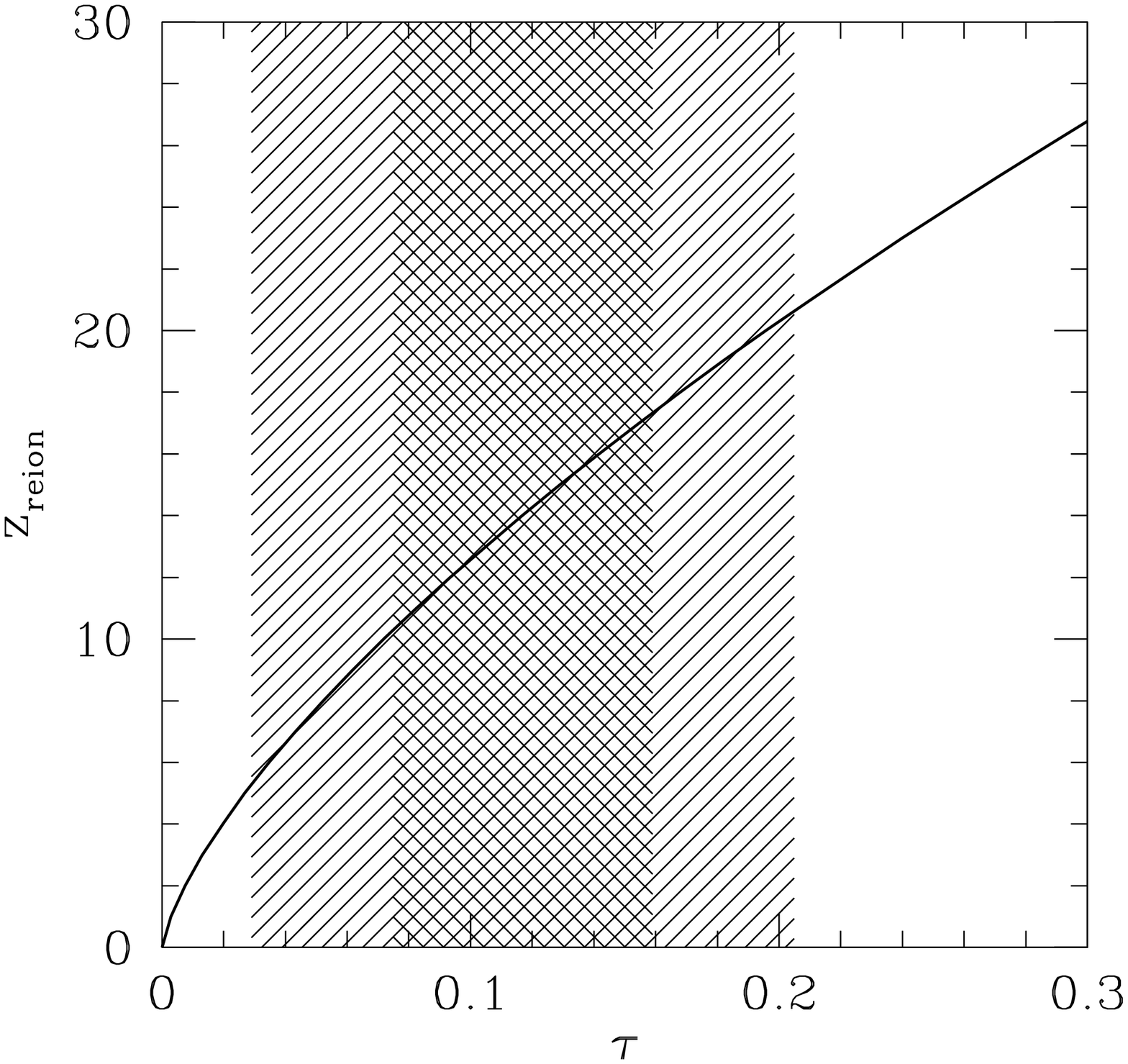}}
\figcaption{\footnotesize The reionization redshift as a function
of optical depth $\tau$ as given by Eq.~18 (solid line). The hatched regions
indicate the 68\% and 95\% confidence limits on $\tau$. This
corresponds to a reionization redshift of $z_{\rm reion}=14^{+7}_{-9}$
(95\% confidence).
\label{tau_z}}
\end{center}}

For the lensing analysis we have selected a sample of galaxies with
$20<R_C<24$. On scales larger than 10 arcminutes the ``B''-mode
vanishes, and therefore we are able to measure the cosmic shear signal
with high accuracy on these scales. On smaller scales we find a small
``B''-mode.  We split the sample in bright $(20<R_C<22)$ and faint
$(22<R_C<24)$ galaxies, and find a significant ``B''-mode for the
bright sample. These galaxies are large compared to the PSF, and
therefore are not very sensitive to errors in the corrections for
observational distortions. However, these galaxies probe only a
relatively small range in redshift, and intrinsic alignments of
galaxies are expected to be important. Intrinsic alignments can
introduce a ``B''-mode and we conclude that the observed signal is
likely to be caused by this effect. A multi-color catalog, which is
currently being created, will provide stronger constraints as it
provides a better redshift discrimination of the galaxies.

The faint sample also shows a small ``B''-mode, but it is smaller
than the ``E''-mode signal. It is not clear what is the cause of this
signal, and hence the correction is uncertain. As a conservative
approach we add the amplitude of the ``B''-mode signal to the uncertainty
in the measurement of the lensing signal. These measurements are then
used to derive constraints on cosmological parameters. 

The relevant cosmological parameters for the analysis presented here
are the mass density $\Omega_m$, the cosmological constant
$\Omega_\Lambda$, the normalization of the power spectrum $\sigma_8$
and the shape parameter $\Gamma$ (note that $\Gamma\sim\Omega_m h$ in
CDM cosmologies). To relate the model predictions to the data we use
photometric redshift distributions for our sample of source galaxies
$(22<R_C<24)$, and marginalize over the uncertainty in the source
redshifts.

We investigated whether the RCS data can be used to constrain
$\Gamma$. Using Gaussian priors for $\Gamma$ (from 2dF) and
$\Omega_m+\Omega_\Lambda$ (from CMB), and the redshift distribution,
we find that we can only place a lower bound on $\Gamma$ for which we
find $\Gamma>0.1+0.16\Omega_m$ (95\% confidence).

We derive joint constraints on $\Omega_m$ and $\sigma_8$, by
marginalizing over $\Gamma$, $\Omega_\Lambda$ and the source redshift
distribution, using different priors. Marginalizing over $\Gamma$ and
$\Omega_\Lambda$, and using a flat prior for the source redshift
distribution, yields a conservative constraint of
$\sigma_8=0.45^{+0.09}_{-0.12}\Omega_m^{-0.50}$ (95\% confidence). 
A better constraint is derived when we use the Gaussian priors, which
yields $\sigma_8=0.46^{+0.05}_{-0.07}\Omega_m^{-0.52}$ (95\% confidence).

Comparison of the RCS results with three other recent cosmic shear
measurements shows excellent agreement (Bacon et al. 2002; Refregier
et al. 2002; van Waerbeke et al. 2002), and we find an ensemble
averaged value of $\sigma_8=0.92\pm0.03$ (68\% confidence). The value
of $\sigma_8$ determined from the RCS data appears to be free of
systematics. It is not clear, however, whether the results from the
other surveys are biased high as a result of residual
systematics. Some recent studies (Reiprich et al. 2001; Seljak et
al. 2001; Viana et al. 2001) of the abundance of rich clusters suggest
low values for $\sigma_8$, which disagree with the lensing results
with high confidence. The reason for this discrepancy is currently
unclear.

The weak lensing results are also in good agreement with CMB
measurements, when we allow the reionization optical depth $\tau$ and
the spectral index $n_s$ to vary. We have presented a simple
demonstration of how the weak lensing results can be used as a prior
for the parameter estimation from CMB measurement.  We found that we
can derive constraints on $\tau$ and $n_s$ (fixing $\Omega_m=0.3$,
$\Omega_\Lambda=0.7$, $\omega_b=0.02$, and $h=0.7$). Doing so, we
obtain $n_s=1.02^{+0.07}_{-0.06}$ and $\tau=0.12\pm0.09$, both with
95\% confidence. We can relate the value of $\tau$ to the reionization
redshift $z_{\rm reion}$ for which we find $z_{\rm
reion}=14^{+7}_{-9}$ with 95\% confidence. We stress, however, that
the results depend on the assumed priors, and that a full combined
analysis of weak lensing and CMB data, marginalizing over all
parameters, is required for definite results. Our preliminary results,
however, are not inconsistent with the latest observations of quasars
at $z\sim 6$, which indicate that the universe may be just exiting the
reionization epoch at that time (Barkana 2002; Becker et al. 2001;
Djorgovski et al. 2001; Fan et al. 2002).

\end{document}